\documentclass[a4paper,11pt]{article}
\pdfoutput=1        

\usepackage{ulem}
\usepackage{jheppub}
\usepackage[T1]{fontenc}
\usepackage{hyphenat}
\usepackage{multicol}
\usepackage{xcolor}
\usepackage{tcolorbox}
\usepackage{slashed}
\usepackage{float}
\usepackage{bbold}
\usepackage{bm}
\usepackage{multirow}
\usepackage[compat=1.1.0]{tikz-feynman}
\usepackage{tikz,pgfplots,pgfplotstable}
\usepackage[utf8x]{inputenc}

\let\emph\textit

\vspace*{3cm}

\title{\boldmath $W^+W^-H$ production through bottom quarks fusion at hadron colliders}

\author[a,b]{Pankaj Agrawal } 
\author[a,b]{and Biswajit Das}

\affiliation[a]{Institute of Physics, \\Sainik School Post, Bhubaneswar 751 005, India}
\affiliation[b]{Homi Bhabha National Institute,\\Training School Complex, Anushakti Nagar, Mumbai 400085, India}

\emailAdd{agrawal@iopb.res.in}
\emailAdd{biswajit.das@iopb.res.in}

\abstract{With the standard model working well in describing the collider data, the focus is now on determining the standard model parameters as well as for any hint of deviation. In particular, the determination of the couplings of the Higgs boson with itself and with other particles of the model is important to better understand the electroweak symmetry breaking sector of the model. In this letter, we look at the process $pp \to WWH$, in particular through the fusion of bottom quarks. Due to the non-negligible coupling of the Higgs boson with the bottom quarks, there is a dependence on the $WWHH$ coupling in this process. This sub-process receives largest contribution when the $W$ bosons are longitudinally polarized. We compute one-loop QCD corrections to various final states with polarized $W$ bosons. We find that the corrections to the final state with the longitudinally polarized $W$ bosons are large. It is shown that the measurement of the polarization of the  $W$ bosons can be used as a tool to probe the $WWHH$ coupling in this process. We also examine the effect of varying $WWHH$ coupling in the $\kappa$-framework.}

\begin{document}

\maketitle

\flushbottom

\newpage

\section{Introduction}
\label{sec:intro}

Standard Model (SM) has been very successful. It has been 
tested in a wide variety of low energy and high energy experiments \cite{conference1,conference2}. Although there is no firmly established conflict between the
data and the standard model predictions, the model is not yet
fully validated. In particular, the Higgs sector  of the model is not yet fully explored. The Higgs potential can
still have many allowed shapes \cite{Agrawal:2019bpm}. Self-couplings of the Higgs boson and its couplings with some of the standard model particles are
still loosely bound. The more precise measurement of the couplings can also lead to hints to beyond the standard model scenarios.

In this letter, we are interested in the coupling of the Higgs boson with the $W$ and $Z$  bosons (Collectively 
referred to as $V$) in particular, we are interested in the quartic $VVHH$ couplings.
In the standard model, the $VVH$ and $VVHH$ couplings are related.
The experimental verification of this relationship is important
to put the standard model on a firm footing. There are scenarios beyond the SM,
where these couplings are either not related or have different relationship \cite{Bishara_2017}.
The ATLAS collaboration has put a bound on this coupling at 
 the Large Hadron Collider (LHC). Using the VBF mechanism of
 a pair of Higgs boson, and using 126 fb$^{-1}$ of data at 13
    TeV, there is a bound of $ -0.43 < \kappa_{V_2H_2} < 2.56$
 at 95$\%$ confidence level \cite{Aad_2021}. Here $ \kappa_{V_2H_2}$ is the scaling factor for the $VVHH$ coupling. However, in this process bound
  on $WWHH$ and $ZZHH$ couplings cannot be separated. The 
    process $p p \to HHV$, where a
   pair of Higgs bosons are produced in association with a $W$ or a $Z$ boson, allows us to separately measure $HHWW$ and $HHZZ$ couplings. Gluon-gluon fusion would contribute to
    $HHZ$ production. This mechanism is important at HE-LHC and FCC-hh.
    However, dependence on the scaling of $HHVV$ coupling is weak. The expected bound
    from the $WHH$ production 
  at the HL-LHC is $ -10.6 < \kappa_{V_2H_2} < 11$ \cite{Nordstrom:2018ceg}, which is quite loose.
  
     Instead of these processes, we consider the process  $pp \to HWW$ at hadron colliders. This process can help us in measuring
    $HHWW$ coupling, independent of $HHZZ$ coupling.
   This processes can take place by both quark-quark and gluon-gluon
   scattering. At a 100 TeV collider, gluon-gluon scattering and bottom-bottom quark
    scattering give important contributions. These contributions depend
    on $HHWW$ coupling. The gluon-gluon contribution is discussed in \cite{Agrawal:2019ffb}. This contribution is significantly lower than the contribution of bottom-bottom scattering. But the contribution of  bottom-bottom scattering is only about $15-20\%$ of the light quarks
    scattering contribution at the 100 TeV center of mass energy (CME) and at the
    leading order (LO), light quarks contribution does not depend
    on $WWHH$ coupling.
    The dependence on this quartic coupling, $WWHH$, can be enhanced if we measure
    the polarization of the final state $W$ bosons.  
    There is significant enhancement of the fraction
    of the bottom-bottom scattering events, when both $W$ bosons in the 
    final states are longitudinally  polarized. The ATLAS and CMS collaborations have measured
    the $W$ polarization at the LHC~\cite{Chatrchyan:2011ig,Aad:2012ky,Aaboud:2019gxl}. We compute the one-loop QCD corrections
    to various combinations of final state $W$ bosons polarization. The longitudinally polarized
    $W$ boson final states receive largest corrections, leading
    to even larger fraction of events with bottom-bottom scattering.  We also scale the $WWHH$ coupling and examine the effect of the NLO QCD corrections and the measurement of the polarization of W bosons. It appears that an analysis of $WWH$ events, when both the $W$ bosons
    are longitudinally polarized, can help in determining the
    $WWHH$ coupling.
    
       The paper is organized as follows. The second and third sections describes the process and the details of the calculations. In the fourth section, we present the
     numerical results, and the last section has the conclusions.

\section{The Process}
\label{sec:prcs}
We are interested in quark-quark scattering  for the production of $WWH$. To study $WWHH$ coupling,
we consider this process in five-flavour scheme.
We study the process $b\:\bar{b}\rightarrow W^+W^-H$ at hadron colliders. 
We take bottom quarks as massless but at the same time, we consider $b\:\bar{b}\:H$ 
Yukawa coupling which is proportional to the mass of the bottom quark. With this consideration, 
the diagrams with $WWHH$ coupling would appear, with the Higgs boson coupling to the bottom 
quark. This coupling would not appear at the leading order (LO) for the other quarks in the initial state.  
This channel has been discussed only with $t\:\bar{t}\:H$ Yukawa couplings \cite{Baglio:2016ofi} but not with $b\:\bar{b}\:H$ Yukawa couplings.

\begin{figure}[!]
\includegraphics [angle=0,width=1\linewidth]{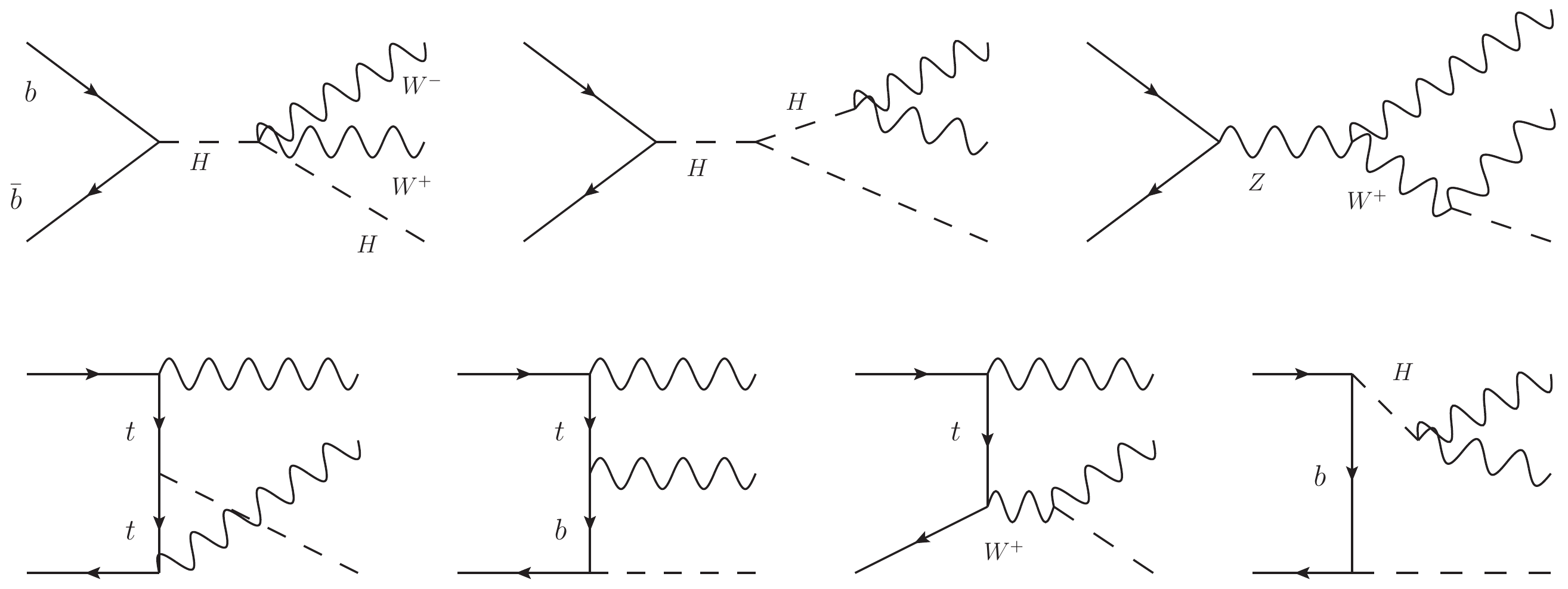}\\
	\caption{A few LO Feynman diagrams for $W^+W^-H$ production in $b\bar{b}$ channel. Diagrams in the upper row are $s$-channel diagrams and in the lower row are $t$-channel diagrams.  }
	\label{fig:bb2wwh_tree_dia}
\end{figure}

At the LO there are 20 diagrams -- 9 s-channel and 11 t-channel. A representative
set of diagrams are displayed in Fig.~\ref{fig:bb2wwh_tree_dia}. Only one of the diagrams has 
$WWHH$ coupling which is one of our main points of interest.
We vary $WWHH$ coupling in order to see its impact on the cross section for the different center of mass energies. There is no strong coupling dependency in the LO diagrams; they solely depend on electroweak couplings. Some of
 the $t$-channel diagrams depend on $t\:\bar{t}\:H$ Yukawa couplings and give large contributions to the LO cross section, due to the top-quark mass dependency of $t\:\bar{t}\:H$ Yukawa coupling.

\begin{figure}[!]
\includegraphics [angle=0,width=1\linewidth]{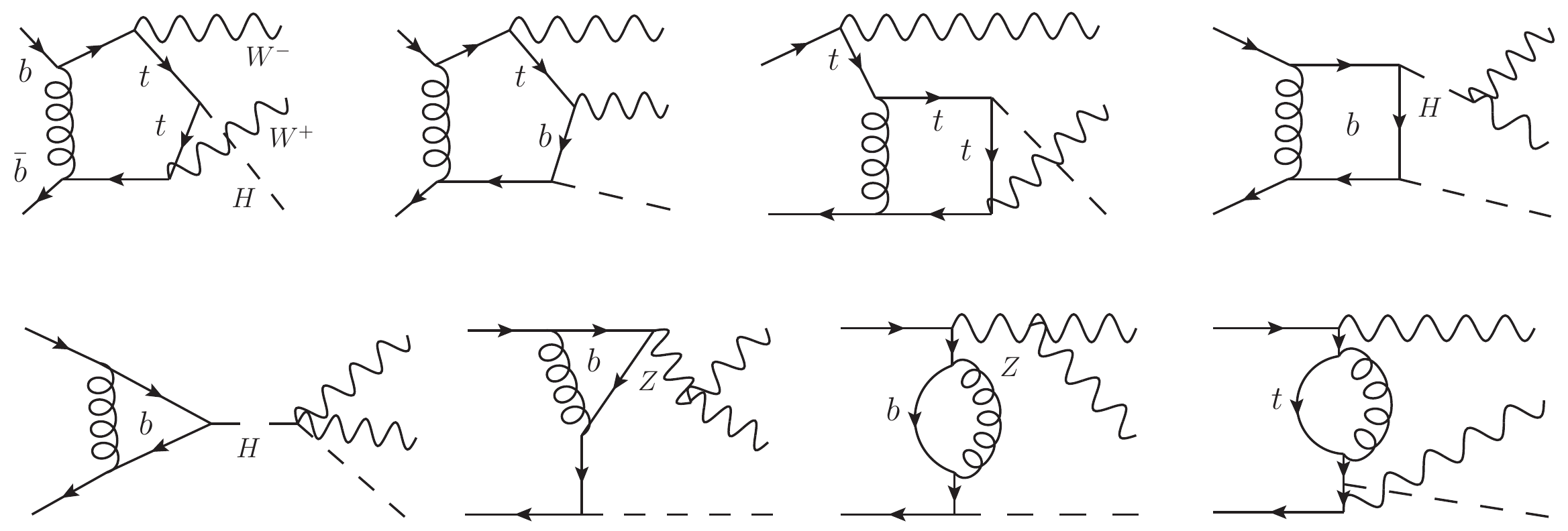}\\
	\caption{Few sample one-loop Feynman diagrams for $W^+W^-H$ production in $b\bar{b}$ channel. }
	\label{fig:bb2wwh_loop_dia}
\end{figure}

\begin{figure}[!]
\includegraphics [angle=0,width=1\linewidth]{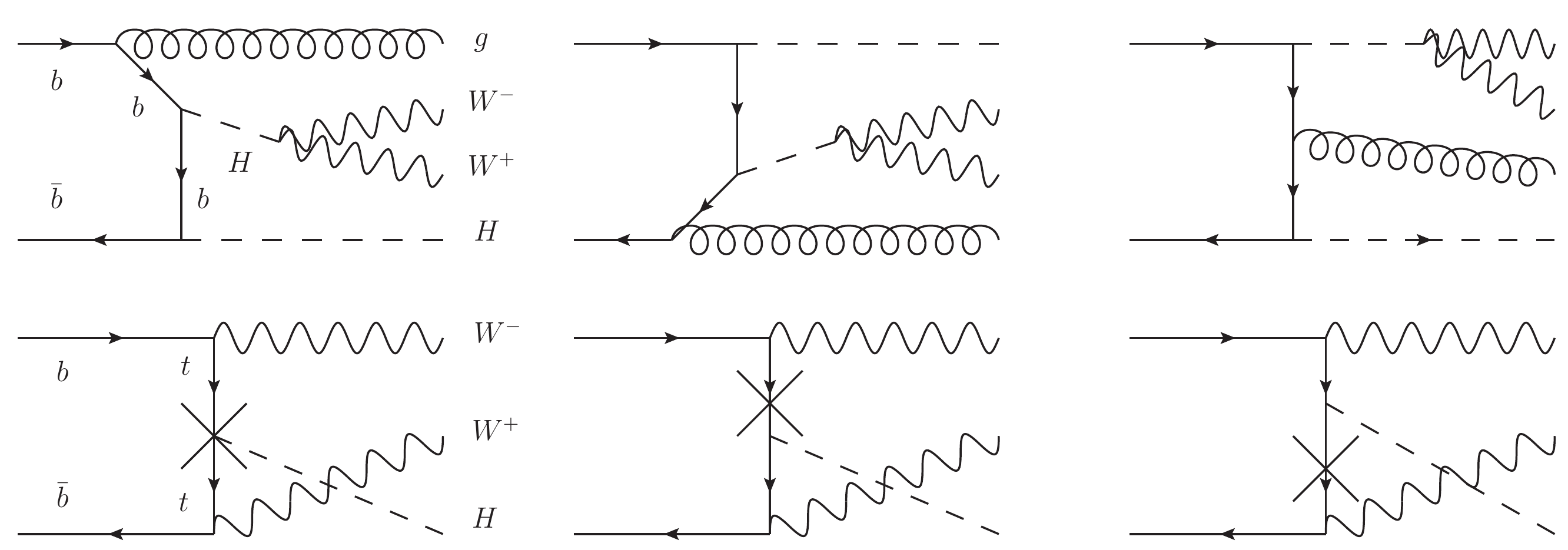}\\
	\caption{Few sample real emission Feynman diagrams for the sub-process $b\:\bar{b}\rightarrow W^+W^-H\:g$ and vertex CT diagrams and self energy CT diagrams for the process $b\:\bar{b}\rightarrow W^+W^-H$. }
	\label{fig:bb2wwh_real_ct_dia}
\end{figure}

To compute the one-loop QCD corrections to this process, we need to include one-loop diagrams
and next-to-leading order (NLO) tree level diagrams. The one-loop diagrams can be categorized as pentagon, box, triangle as well as bubble diagrams. There are $3$ pentagon diagrams, $14$ box diagrams, $34$ triangle diagrams, and $14$ bubble diagrams.  A few representative  NLO diagrams are displayed in Fig.~\ref{fig:bb2wwh_loop_dia}. There is only one one-loop diagram (triangle) which has $WWHH$ coupling. Bubble diagrams are UV divergent and a few triangle diagrams are also UV divergent. To remove UV divergence from the amplitude, counterterm (CT) diagrams need to be added to the virtual amplitudes. There are $15$ vertex CT diagrams and $14$ self energy CT diagrams.  A set of CT diagrams are shown in Fig.~\ref{fig:bb2wwh_real_ct_dia}. Also, most of the virtual diagrams are infrared (IR) singular. In order to remove IR singularities from the virtual diagrams, one needs to include real emission diagrams. There are three such processes. These processes are a) $b\bar{b}\rightarrow W^+W^-Hg$, b) $g\bar{b}\rightarrow W^+W^-H\bar{b}$ and c) $bg\rightarrow W^+W^-Hb$. There are $54$ Feynman diagrams for the each of these processes. We have shown a few diagrams for the first sub-process in Fig.~\ref{fig:bb2wwh_real_ct_dia}. All these diagrams have been generated using a {\tt Mathematica} package, {\tt FeynArts} \cite{Hahn:2000kx}.

\section{Calculations and Checks}
\label{sec:calc_check}

 We have to perform $2 \to 3$ and $2 \to 4$ tree level and $2 \to 3$
 one loop calculations. For the calculation we use helicity methods.
 As a starting point, we consider a few prototype diagrams in each case. With suitable
 crossing, and coupling choices, we can compute rest of the diagrams.
  We compute helicity amplitudes at the matrix element level for the
  prototype diagrams. These helicity amplitudes can be used to probe the physical observables dependent on the polarization of external particles. As mentioned before, the $b$-quarks are treated as massless quarks because of their small mass and we use massless spinors for $b$-quarks. 
  The tree level helicity amplitudes can be written in terms of the spinor products $\langle pq\rangle$ and $[pq]$ \cite{Peskin:2011in}. For one-loop amplitudes, we use an extra object - the vector  current $\langle p \gamma^\mu q]$.  We take the functional form of the spinor products $\langle pq\rangle$ or $[pq]$ from Ref. \cite{Kleiss:1986qc} and we extend their treatment to calculate the functional form of the vector  current $\langle p \gamma^\mu q]$. We have checked that the calculated $\langle p \gamma^\mu q]$ satisfies various spinor identities. We adopt four-dimensional-helicity (FDH) scheme \cite{BERN1992451,Gnendiger:2017pys} to compute the amplitudes. In this scheme all spinors, $\gamma$-matrices algebra are computed in $4$-dimensions. We use package {\tt FORM} \cite{Vermaseren:2000nd} to implement the helicity formalism.  

Using {\tt FORM}, we write helicity amplitude in terms of spinorial objects, scalar products of momenta and polarizations. For the one-loop calculations,
we also have tensor and scalar integrals. The one-loop scalar integrals are computed using the package OneLoop \cite{vanHameren:2010cp}. We use an in-house reduction code, OVReduce \cite{Agrawal:2012df,Agrawal:1998ch}, to compute tensor integrals in dimensional regularization. Finally, the phase space integrals have been done with the advanced Monte-Carlo integration ({\tt AMCI}) package \cite{Veseli:1997hr}. In {\tt AMCI}, the {\tt VEGAS}\cite{Lepage:1977sw} algorithm is implemented using parallel virtual machine ({\tt PVM}) package \cite{10.7551/mitpress/5712.001.0001}.

Few checks have been performed to validate the amplitudes. The one-loop
amplitudes have both ultraviolet (UV) and infrared (IR) singularities. 
UV singularities are removed by using counter-term (CT) diagrams, and
the IR divergences are removed using Catani-Seymour (CS) dipole substraction
methods. Cancellation of these diverges are powerful checks on the
calculation.
All UV singularities are removed by fermionic mass and wave function renormalization. There are no UV singularities coming from pentagon and box diagrams as there are no 4-point box tensors in those amplitudes. UV singularities are coming from the triangle as well as bubble diagrams. The appropriate vertex and self energy counterterms (CT) diagrams have been added in total amplitude which gives renormalized amplitude. A few sample CT diagrams are depicted in Fig. 4. We use the $\rm \overline{MS}$ scheme for massless fermions and the on-shell subtraction scheme for massive fermions.

The next check is infrared (IR) singularity cancellation. We implement the Catani-Seymour dipole subtraction method \cite{Catani:1996vz} for the cancellation of IR singularities. Except bubble diagrams, all other virtual diagrams are IR singular. Collectively all IR singularities coming from virtual diagrams cancel with IR singularities coming from real emission diagrams. 

Following the Catani-Seymour method, the NLO cross section can be written as  
\begin{eqnarray}
\sigma^{NLO} &=& \int_{m+1} d\sigma^R +\int_m d\sigma^V \nonumber \\
&=& \int_{m+1}(d\sigma^R-d\sigma^A)+\int_m (d\sigma^V+\int_1d\sigma^A) \;.\label{eq:dp_VA}
\end{eqnarray}
Where $d\sigma^R$, $d\sigma^V$ and $d\sigma^A$ are exclusive cross section, one-loop virtual correction and approximation term respectively. $d\sigma^A$ has the same pointwise singular behaviour as $d\sigma^R$ and hence behaves as a local counterterm for $d\sigma^R$ and then first integration can be performed safely in $\epsilon\rightarrow0$ limits. The second term of the second integral will give dipole {\textbf{\textit I}} term which will remove all the infrared singularities from virtual correction and add a finite contribution. 
The dipole $\textit{\textbf{I}}$ factor comes from analytical integration of $d\sigma^A$ in $d$-dimensions over one-parton phase space. It can be written as
\begin{eqnarray}
\int_1 d\sigma^A = d\sigma^B\otimes\textit{\textbf{I}} 
\end{eqnarray}  
Where $d\sigma^B$ is born level cross section and the symbol $\otimes$ describes phase space convolution and sum over spin and color indices. The term $d\sigma^B\otimes\textit{\textbf{I}}$ is evaluated over the rest of $m$-parton phase space and cancels all singularities from renormalized virtual amplitudes. As discuss before, we use the FDH scheme, so we take $\textit{\textbf{I}}$ term in the FDH scheme. The term $\textit{\textbf{I}}$ given in Ref. \cite{Catani:1996vz} is in conventional dimensional regularization (CDR) scheme and in any other regularization scheme (RS) it is given as \cite{Catani:1996pk}
\begin{eqnarray}
\textit{\textbf{I}}^{\:\rm RS}( \{p\},\epsilon) = \textit{\textbf{I}}^{\:\rm CDR}(\{p\},\epsilon) -\frac{\alpha_s}{2\pi}\sum_I\widetilde{\gamma}_I^{\:\rm RS} + \mathcal{O}(\epsilon)\:.
\end{eqnarray}
In the FDH scheme, $\widetilde{\gamma}_I^{\:\rm RS}$ are 
\begin{eqnarray}
\widetilde{\gamma}_q^{\:\rm FDH} = \widetilde{\gamma}_{\bar{q}}^{\:\rm FDH}=\frac{1}{2}C_F\:,\quad \widetilde{\gamma}_g^{\:\rm FDH}=\frac{1}{6}C_F\:.
\end{eqnarray} 
Now with this $\textit{\textbf{I}}$ term, we have checked that the integration in the second term of Eq.~\ref{eq:dp_VA} is IR safe. Also, there are other terms in the dipole subtraction method, called $\textit{\textbf{P}}$ and $\textit{\textbf{K}}$ terms which will add finite contributions to $\sigma^{NLO}$. These terms come from the factorization of initial-state singularities into parton distribution functions. The color operator algebra, explicit form of ${ \textit{\textbf{V}}_{ij,k}}$, $\textit{\textbf{I}}$, $\textit{\textbf{P}}$ and $\textit{\textbf{K}}$ are given in Ref.~\cite{Catani:1996vz}.

There are three real emission sub-processes that can contribute to $\sigma^{NLO}$. These processes are 
\begin{equation}
a)\quad b\:\bar{b}\rightarrow W^+W^-H\:g\quad
b)\quad g\:\bar{b}\rightarrow W^+W^-H\:\bar{b}\quad
 c) \quad b\:g\rightarrow W^+W^-H\:b\:,
\end{equation}
as these processes mimic the Born level process in soft and collinear regions. Due to large contributions, top resonance in the last two processes jeopardizes the perturbative calculation. The cross sections for these two processes are five to six times higher than the Born level cross section. One can't remove those top resonant diagrams as it will affect the gauge invariance and we have checked that the interference between resonant and non-resonant diagrams coming from the off-shell region is large which will again ruin the perturbative computations. There are several techniques to remove these on-shell contributions safely \cite{Grazzini:2016ctr,Denner:2012yc,Cascioli:2013wga,Gehrmann:2014fva}. One can also restrict resonant top momenta out of the on-shell region and can have contribution only from the off-shell region. To implement the last technique with a standard jet veto, one needs a very large number of phase space points to get a stable cross section. The implementation of these techniques is beyond the scope of this paper. Instead of these techniques, we exclude the last two channels by assuming $b$-quark tagging with $100\%$ efficiency \cite{Bierweiler:2012kw,Baglio:2016ofi}.
\section{Numerical Results}
\label{sec:numr_res}
The sub-process $b\:\bar{b}\rightarrow W^+W^-H$ gives a significant contribution to the main process $p\:p\rightarrow W^+W^-H$. We calculate the NLO QCD contribution to this process. In particular we focus on 
the corrections to  cross sections and distributions for various polarization configurations of the final state particles. We also probe variation of cross sections with $WWHH$ anomalous coupling.
Some of the Feynman diagrams, tree-level diagrams, as well as one-loop diagrams are heavy vector bosons, Higgs boson and top quark mediated. We use complex-mass scheme (CMS) \cite{DENNER200622} throughout our calculation to handel the resonance instabilities coming from these massive unstable particles. We take Weinberg angle as $cos^2\theta=(m_W^2 - i\Gamma_Wm_W)/(m_Z^2 - i\Gamma_Zm_Z)$. The input SM parameters are \cite{GRAZZINI2020135399}: $m_W = 80.385$GeV, $\Gamma_W=2.0854$GeV, $m_Z = 91.1876$GeV, $\Gamma_Z=2.4952$GeV, $m_H=125$GeV, $\Gamma_H=0.00407$GeV, $m_b=4.92$GeV, $m_t=173.2$GeV, $\Gamma_t = 1.44262$GeV. There are several pieces in the one-loop calculation which contribute to total $\sigma^{NLO}$. As we have discussed above, virtual amplitudes, CT amplitudes, dipole $\textit{\textbf{I}}$, $\textit{\textbf{P}}$ and $\textit{\textbf{K}}$ terms, dipole subtracted real emission amplitudes contribute to the finite part. We find that there are significant contributions from all these pieces except dipole subtracted real emission amplitudes which gives an almost vanishing contribution.

We use {\tt CT14llo}  and {\tt CT14nlo} PDF sets \cite{Dulat:2015mca} for LO ($\sigma^{LO}$) and NLO ($\sigma^{NLO}$) cross sections calculation. We use these PDF sets through {\tt LHAPDF} \cite{Whalley:2005nh} libraries. As mentioned before we calculate the cross sections in three different CMEs corresponding to current and proposed future colliders. We choose renormalization ($\mu_R$) and factorization ($\mu_F$) scales dynamically as 
\begin{equation}
\mu_R=\mu_F=\mu_0=\frac{1}{3}\Big(\sqrt{p_{T,W^+}^2+M^2_{W}}+\sqrt{p_{T,W^-}^2+M^2_{W}}+\sqrt{p_{T,H}^2+M^2_{H}}\Big),\label{eq:scale_mu}
\end{equation} 
where $p_{T,W}$, $p_{T,H}$ are the transverse momenta and $M_W$, $M_H$ are the masses of $W$ and Higgs bosons. We measure the scale uncertainties by varying both $\mu_R$ and $\mu_F$ independently by a factor of two around the $\mu_0$ given in Eq. \ref{eq:scale_mu}.
\subsection{Results for the SM}
\label{subsec:numr_res_sm}
We have listed the cross sections for different CMEs with their respective scale uncertainties in Table~\ref{table:bb2WWH_qcd}. As we see in Table~\ref{table:bb2WWH_qcd} the LO cross sections are $252$, $1313$ and $20671$ {\it ab} whereas NLO cross sections are $300$, $1653$ and $27221$ {\it ab} at $14$, $27$ and $100$ TeV CMEs respectively. The cross section rapidly increases with CME as PDFs for $b$-quarks are small for lower energies. The relative enhancements \big(RE $=\frac{\sigma^{\rm NLO}_{\rm QCD}-\sigma^{\rm LO}}{\sigma^{\rm LO}}$\big) due to NLO QCD correction are also presented in that table. The RE also increases with CME and it is $19.0\%$, $25.9\%$ and $31.7\%$ for $14$, $27$ and $100$ TeV CMEs respectively. We have calculated scale uncertainty as the relative change in the cross sections for the different choices of scales within bound $0.5 \mu_0\leq\mu_R/\mu_F\leq 2\mu_0$. We see that the NLO uncertainties are a little bit higher than the LO. As there is no strong coupling ($\alpha_s$) at the Born level, the LO uncertainties come largely from 
the factorization scale whereas at the NLO the uncertainties come from both, factorization as well as renormalization scales. To see the different scale uncertainties separately, we vary $\mu_R$ and $\mu_F$ independently. We see the renormalization scale uncertainty varies from $\sim -10\%$ to $\sim3\%$ and the factorization scale varies from $\sim-14\%$ to $\sim 18\%$ at NLO depending on CMEs from $14$ to $100$ TeV.

\begin{table}[H]
\begin{center}
\begin{tabular}{|c|c|c|c|} 
\hline
 CME(TeV)& $\sigma^{LO}$[ab] & $\sigma^{NLO}_{QCD}$[ab]&RE\\ 
\hline
 $14$ & $252^{+14.7\%}_{-17.5\%}$&$300^{+18.0\%}_{-20.7\%}$ &$19.0\%$\\ 
$27$& $1313^{+17.4\%}_{-19.3\%}$ & $1653^{+19.0\%}_{-20.6\%}$&$25.9\%$\\ 
$100$&  $20671^{+20.3\%}_{-21.0\%}$& $27221^{+20.0\%}_{-21.1\%}$&$31.7\%$ \\ 
\hline
\end{tabular}
\caption{The LO and NLO cross sections for different collider CMEs with their respective scale uncertainties. RE is the relative enhancement of the total cross section from the Born level cross section.}
\label{table:bb2WWH_qcd}
\end{center}
\end{table}

As discussed before, we probe the contributions from different polarization configurations of the final state $W$ bosons to the LO and NLO cross sections. The right-handed, left-handed and longitudinal polarization of a $W$ boson are denoted as `+', `-', and `0'.
 The contributions of different nine polarization combinations of final state $W$ bosons are given in Table~\ref{table:bb2WWH_pol} for $14$, $27$ and $100$ TeV CMEs.  We see that the large contributions are coming from the longitudinal polarization states and among them, the `00' combination gives the largest contribution to the total cross sections.  Relative enhancement (RE) for the `00' combination increases with the CME and it becomes $\sim75\%$ at $100$ TeV. In the $R_{\xi}$ gauge, the pseudo Goldstone bosons couple to massive fermions with a coupling proportional to the 
 mass of the fermion. These pseudo Goldstone boson 
 represents the longitudinal polarization state of
 a $W$ boson. This leads to larger values of the cross section  in longitudinal polarization combinations due to heavy fermion mediated diagrams. These longitudinal polarization modes are useful for background suppression to this process. The background may come from the processes with gauge bosons or gluons or photons couplings with light quarks. The negligible masses of the light quarks ($u,d,s$ and $c$) lead to the suppression of backgrounds in polarization combinations that includes longitudinal polarization. 
\begin{table}
\begin{center}

\begin{tabular}{|c|c|c|c|c|c|c|c|c|c|}
\hline
\multirow{1}{*}{Pol.} & \multicolumn{3}{c|}{$14$\:TeV\:} & %
    \multicolumn{3}{c|}{$27$\:TeV\:}& %
    \multicolumn{3}{c|}{$100$\:TeV\:}\\
\cline{2-10}
\multirow{1}{*}{com.} & $\sigma^{ LO}$ & $\sigma^{NLO}_{QCD}$ & RE(\%) &$\sigma^{LO}$ & $\sigma^{NLO}_{QCD}$& RE(\%)&$\sigma^{ LO}$ & $\sigma^{NLO}_{QCD}$ & RE(\%)\\
\hline
 $++$ &$15$&$18$&$20.0$&$70$&$88$&$25.7$&$841$&$1071$&$27.3$ \\
$+-$&$20$& $25$ &$25.0$&$95$&$127$&$33.7$ &$1144$&$1503$&$31.2$\\
$+0$&$42$ & $50$&$19.0$&$214$& $274$& $28.0$&$3105$&$3436$&$10.7$\\
$-+$& $5$& $6$&$20.0$&$22$& $28$ &$27.2$&$276$&$341$&$23.6$\\
$--$&$15$ & $18$&$20.0$&$71$& $89$ &$25.4$&$844$&$1056$&$25.1$\\
$-0$&$25$&  $28$&$12.0$&$126$& $149$ &$18.3$&$1750$&$1411$&$-19.4$\\
$0+$& $25$& $28$&$12.0$&$127$& $149$&$17.3$ &$1783$&$1366$&$-23.4$\\
$0-$&$41$& $50$&$22.0$&$216$& $275$&$27.3$ &$3103$&$3310$&$6.7$\\
$00$&$64$&  $75$&$17.2$&$372$& $468$ &$25.8$&$7760$&$13610$&$75.4$\\
\hline
$\sum$&$252$&  $300$&$19.0$&$1313$& $1653$ &$25.9$&$20671$&$27221$&$31.7$\\
\hline

\end{tabular}
\caption{The LO and NLO cross sections and their relative enhancements (RE) for different polarization combinations of final state $W$ bosons and their sum at $14$, $27$ and $100$ TeV CMEs. The results are in {\it ab} unit.}
\label{table:bb2WWH_pol}
\end{center}
\end{table}

To find the relative contribution of the bottom-bottom scattering to the $pp\rightarrow W^+W^-H$ process, we compute the cross sections in other $q\bar{q}$ channels along with the $b\bar{b}$ channel. The results are presented in Table~\ref{table:qq2WWH_qcd}. The cross sections in $q\bar{q}$ channels ($4$FNS) have been calculated using {\tt MagGraph5\_aMC5@NLO} \cite{Alwall:2014hca}. {\tt MagGraph5\_aMC5@NLO} cannot compute the one-loop QCD corrections to the $b\bar{b}$ channel due to the presence of the
resonances in the diagrams. As we see in Table~\ref{table:qq2WWH_qcd}, the $b\bar{b}$ channel gives significant contributions to the full process $pp\rightarrow W^+W^-H$. The $b\bar{b}$ channel contributes $\sim2.6\%$ to the LO and $\sim2.3\%$ to the NLO cross sections  at $14$ TeV and  $\sim17.6\%$ to the LO and $\sim14.7\%$ to the NLO  cross sections at $100$ TeV of process $pp\rightarrow W^+W^-H$. These numbers are calculated without the channels $gg\rightarrow W^+W^-H$, which can also add a significant contribution to the process $pp\rightarrow W^+W^-H$~\cite{Agrawal:2019ffb,Baglio:2016ofi}. If one adds $gg$ channel, these numbers will be changed accordingly. As we see in Table~\ref{table:qq2WWH_qcd}, the corrections are pretty high in $q\bar{q}$ channels ($4$FNS). In those channels, {\tt MadGraph5\_aMC@NLO} includes all real emission diagrams and the results are complete but we impose jet veto on $b$-quarks with $100\%$ efficiency for real emission diagrams to overcome certain difficulties discussed in Sec. \ref{sec:calc_check}. The proper inclusion of all real emission diagrams may increase the QCD correction significantly in $b\bar{b}$ channel.  

\begin{table}[H]
\begin{center}

\begin{tabular}{|c|c|c|c|c|}
\cline{1-5}
{ \multirow{2}{*}{channel}}& \multicolumn{2}{c|}{$14$\:TeV\:} & %
    \multicolumn{2}{c|}{$100$\:TeV\:}\\
\cline{2-5}
 & $\sigma^{LO}$[ab] & $\sigma^{NLO}_{QCD}$[ab] &$\sigma^{LO}$[ab] & $\sigma^{NLO}_{QCD}$[ab] \\
\hline 
$4$FNS& $9564$&$13290$&$117500$&$184600$ \\
\hline
$b\bar{b}$& $252$&$300$&$20671$&$27221$ \\
\hline
\end{tabular}
\end{center}
\caption{The LO and NLO cross sections in $4$FNS and the $b {\bar b}$ channel at $14$ and $100$ TeV CMEs. The results for $4$FNS have been obtained using {\tt MagGraph5\_aMC5@NLO}. The $b\bar{b}$ channel results are from our code.  }
\label{table:qq2WWH_qcd}
\end{table}

\begin{figure}[!hbt]
\includegraphics [angle=0,width=0.5\linewidth]{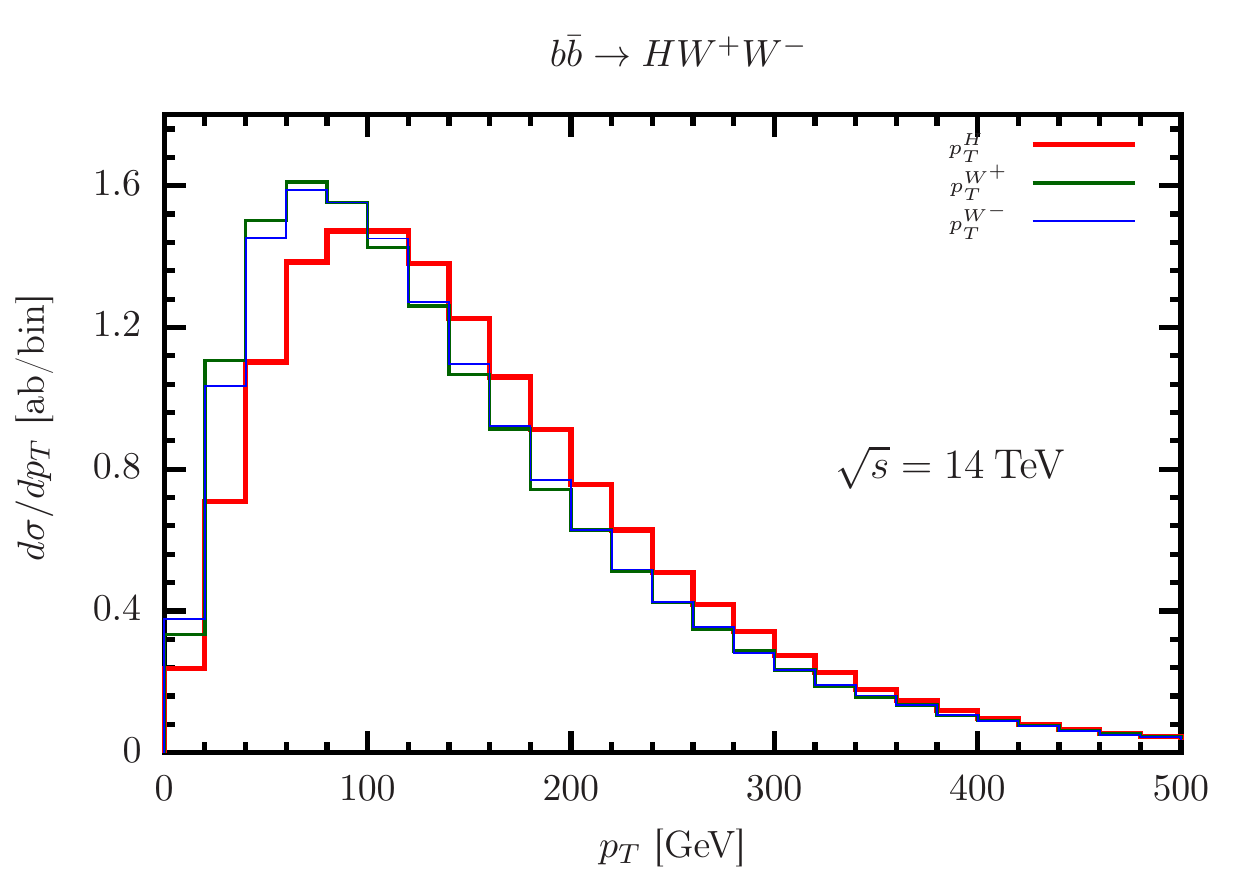}
\includegraphics [angle=0,width=0.5\linewidth]{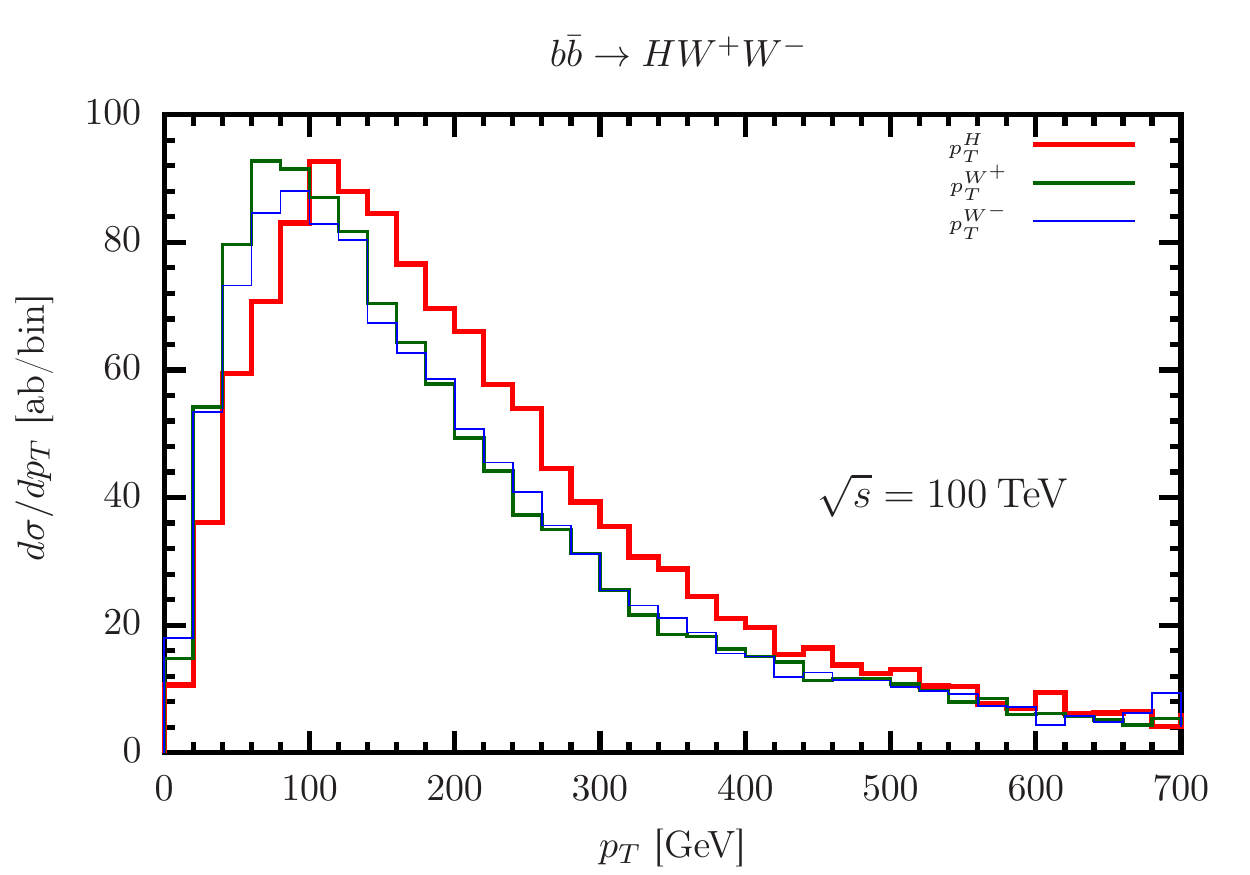}
\includegraphics [angle=0,width=0.5\linewidth]{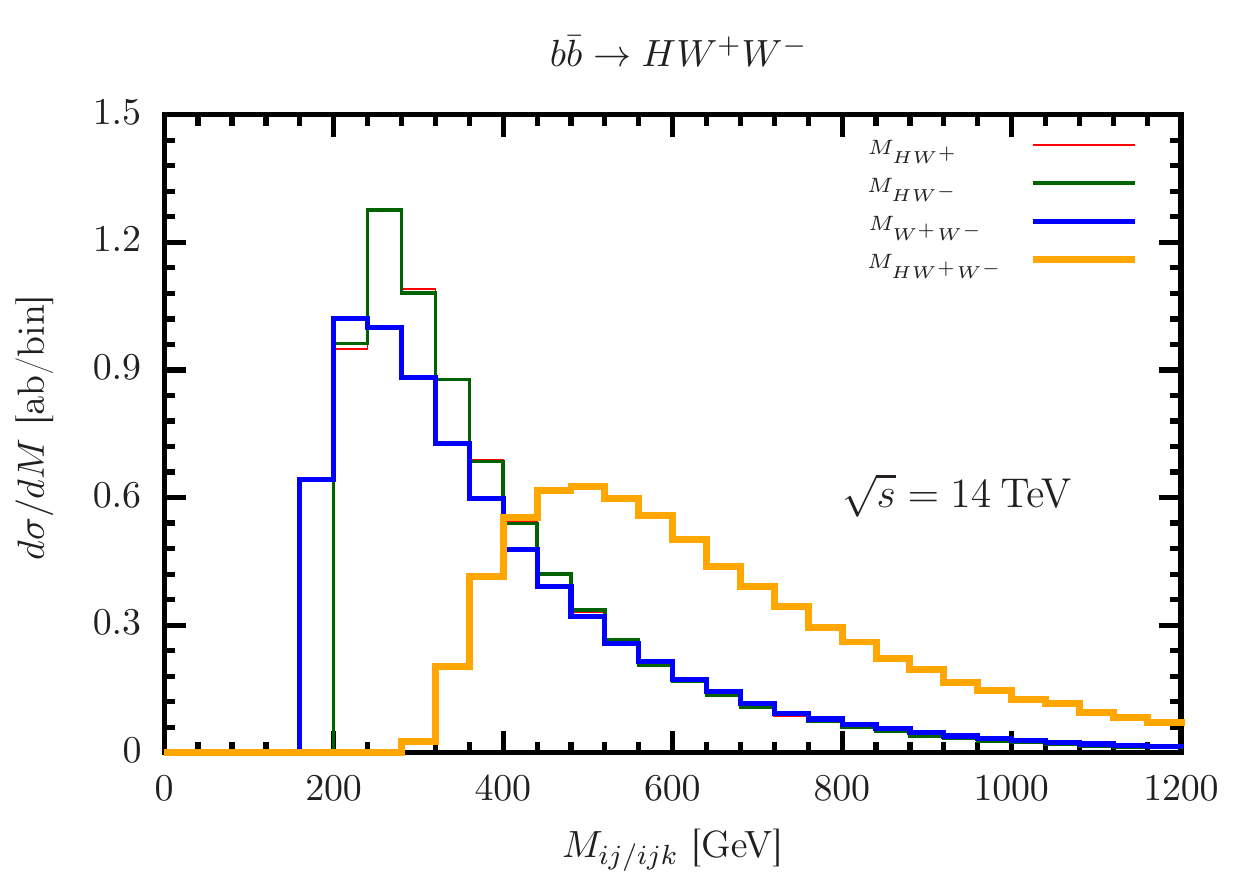}
\includegraphics [angle=0,width=0.5\linewidth]{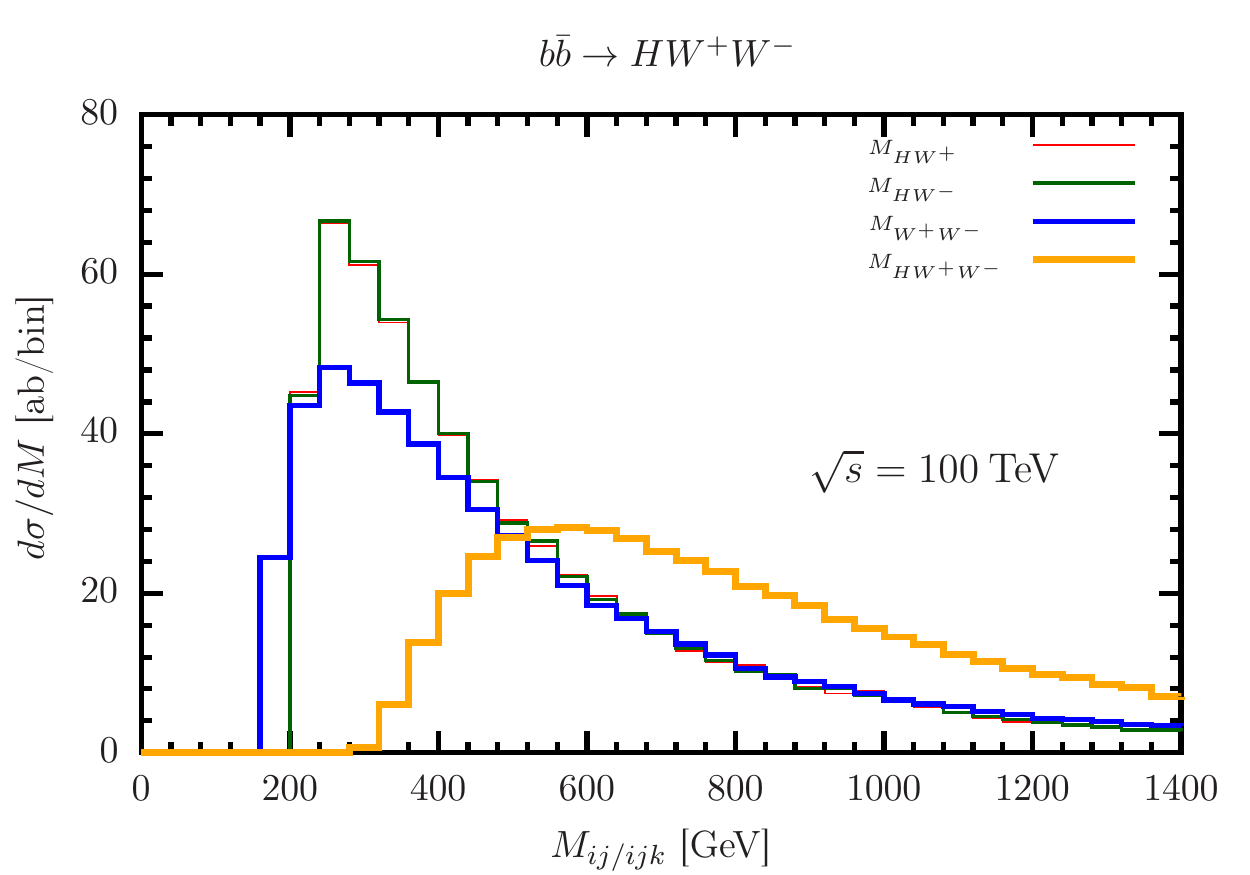}
\caption{ The NLO differential cross section distribution with respect to transverse momentum ($p_T$) and invariant masses ($M_{ij/ijk}$) for $14$ and $100$ TeV CMEs.}
\label{fig:bb2wwh_NLO_pt_inv_mas}
\end{figure}

\begin{figure}[!hbt]
\includegraphics [angle=0,width=0.5\linewidth]{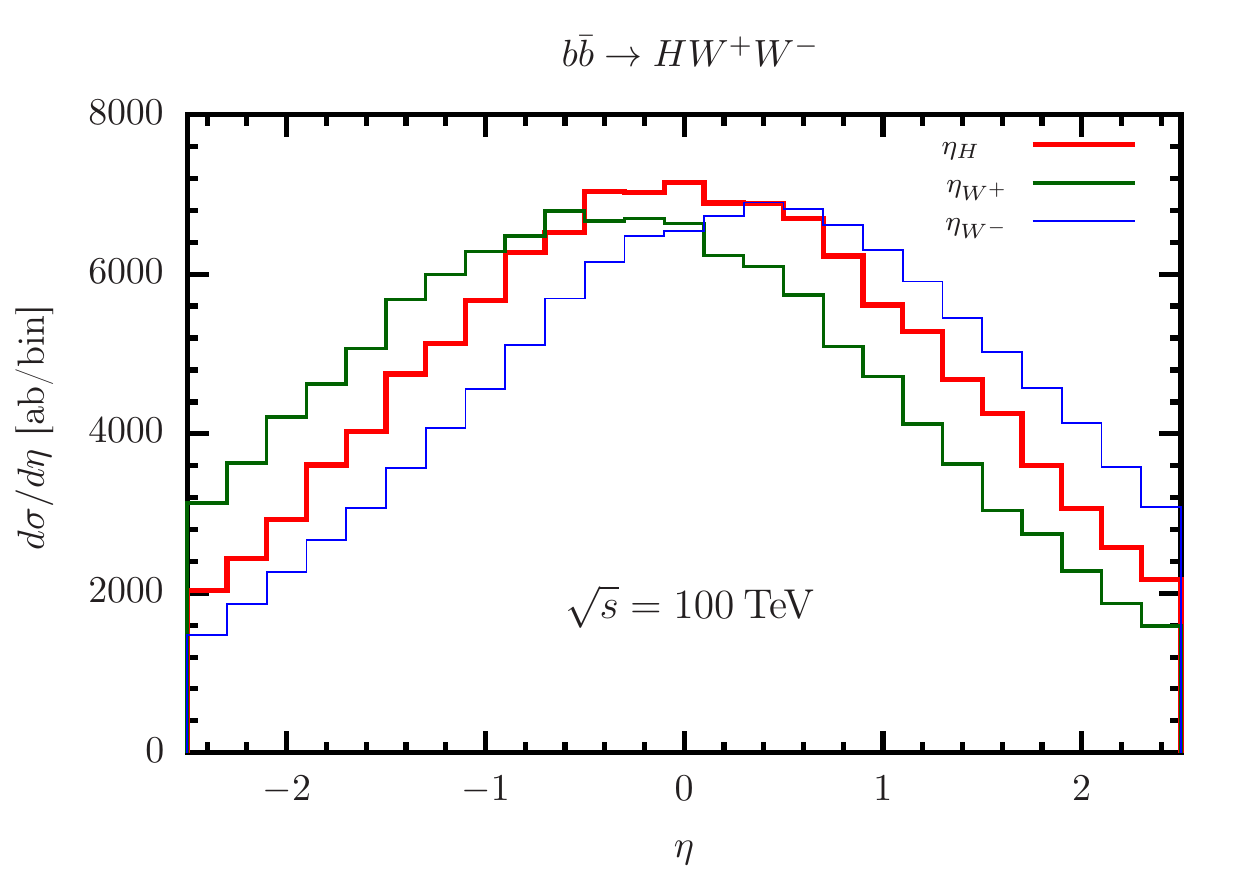}
\includegraphics [angle=0,width=0.5\linewidth]{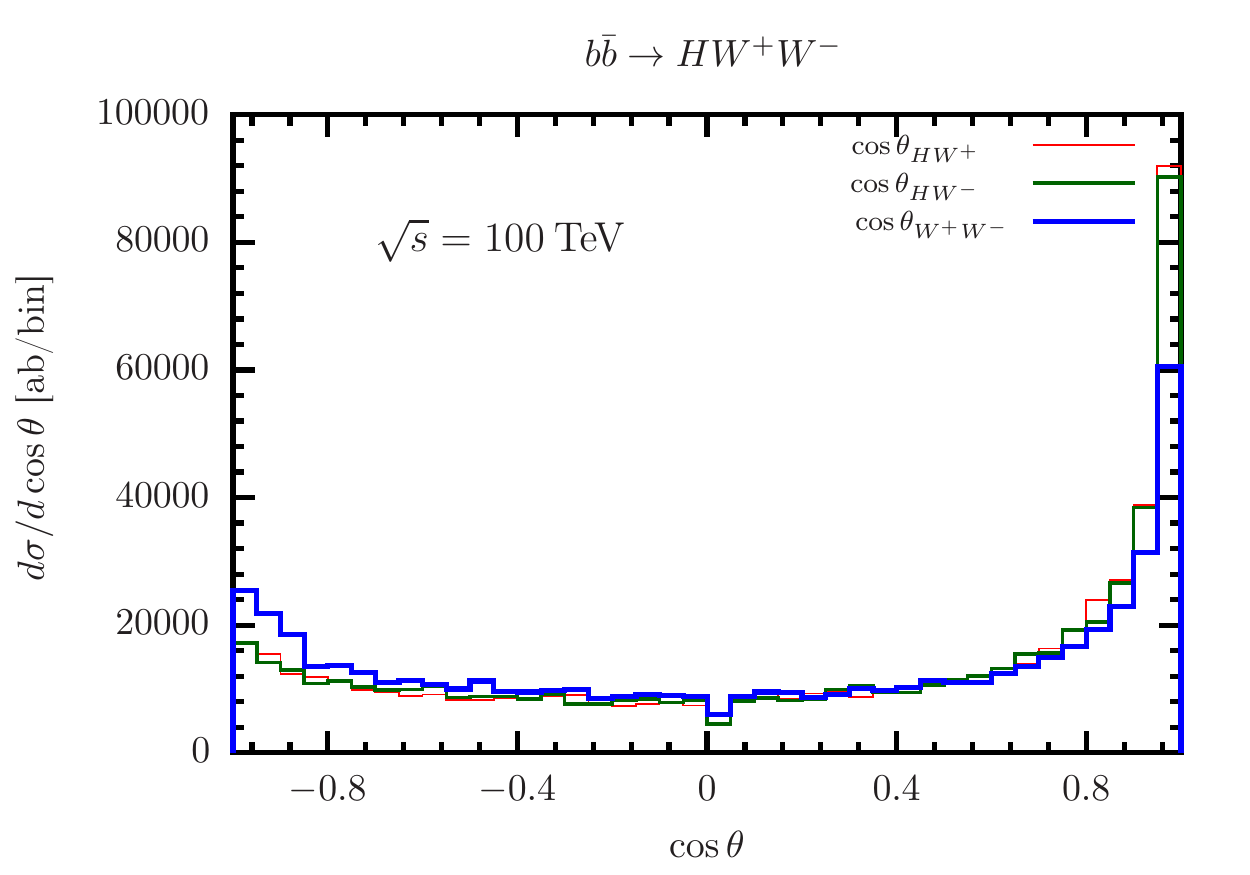}
\caption{ The NLO differential cross section distribution with respect to rapidity ($\eta$) and cosine angle ($\cos\theta$) between the two final state particles at $100$ TeV CME. Plots for 14 TeV are similar.} 
\label{fig:bb2wwh_NLO_eta_cos0}
\end{figure}

\begin{figure}[!hbt]
\includegraphics [angle=0,width=0.5\linewidth]{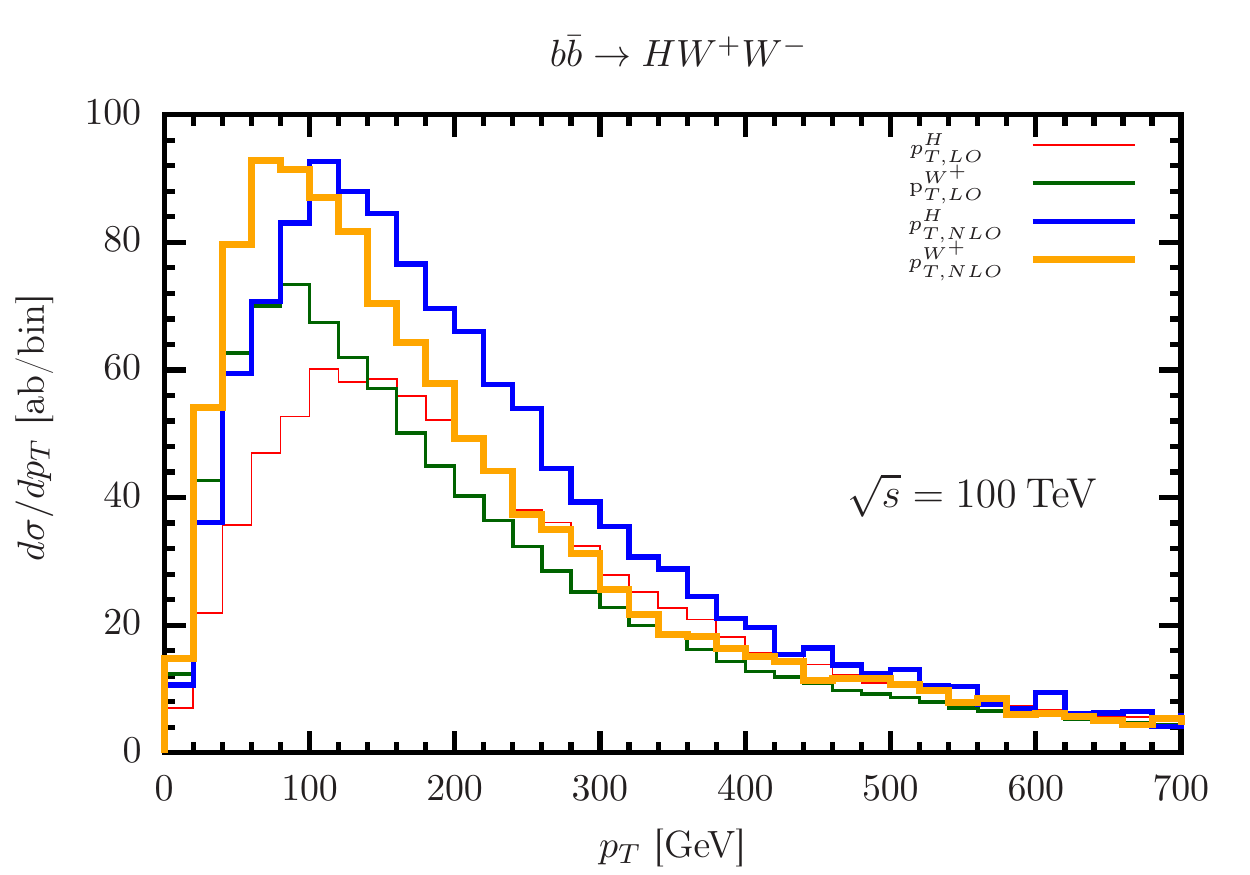}
\includegraphics [angle=0,width=0.5\linewidth]{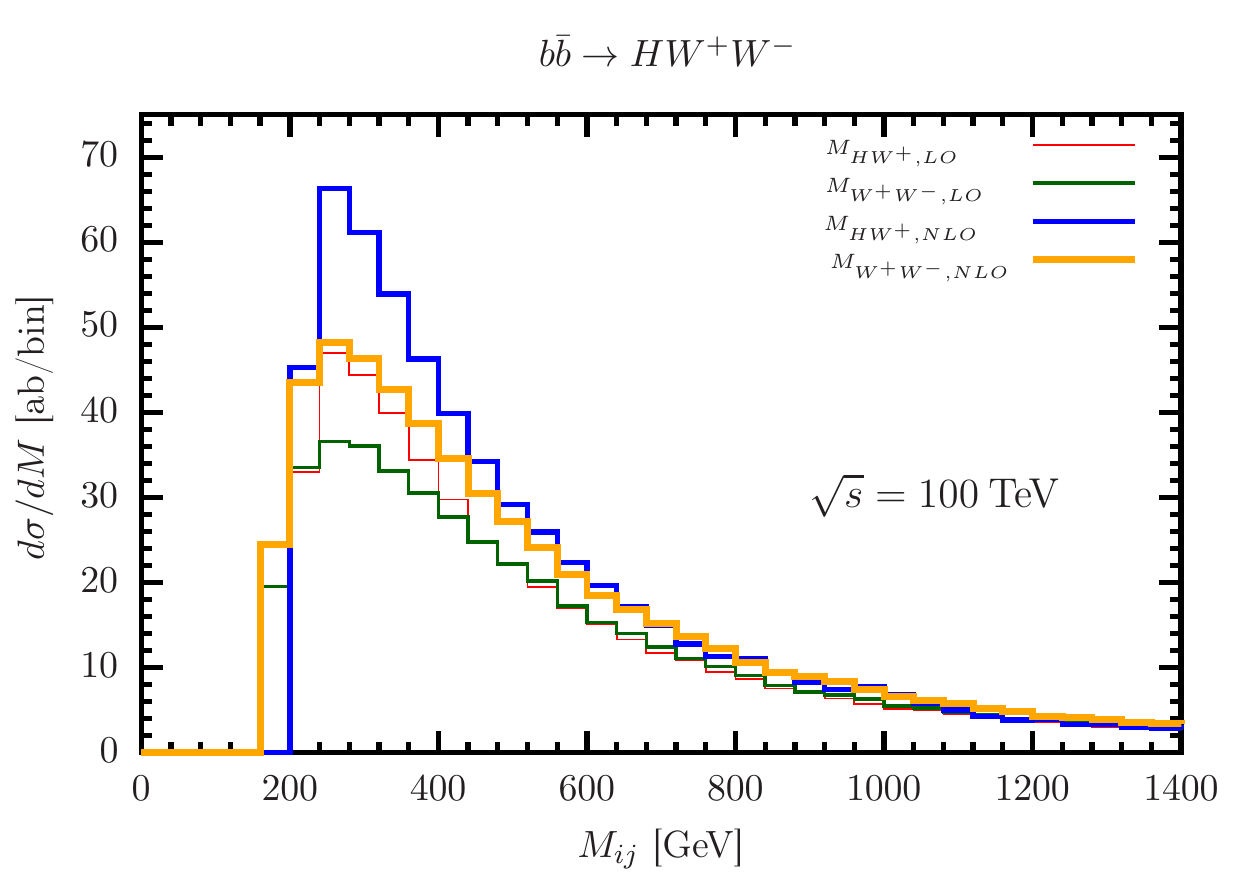}
	\hspace*{4cm} \includegraphics [angle=0,width=0.5\linewidth]{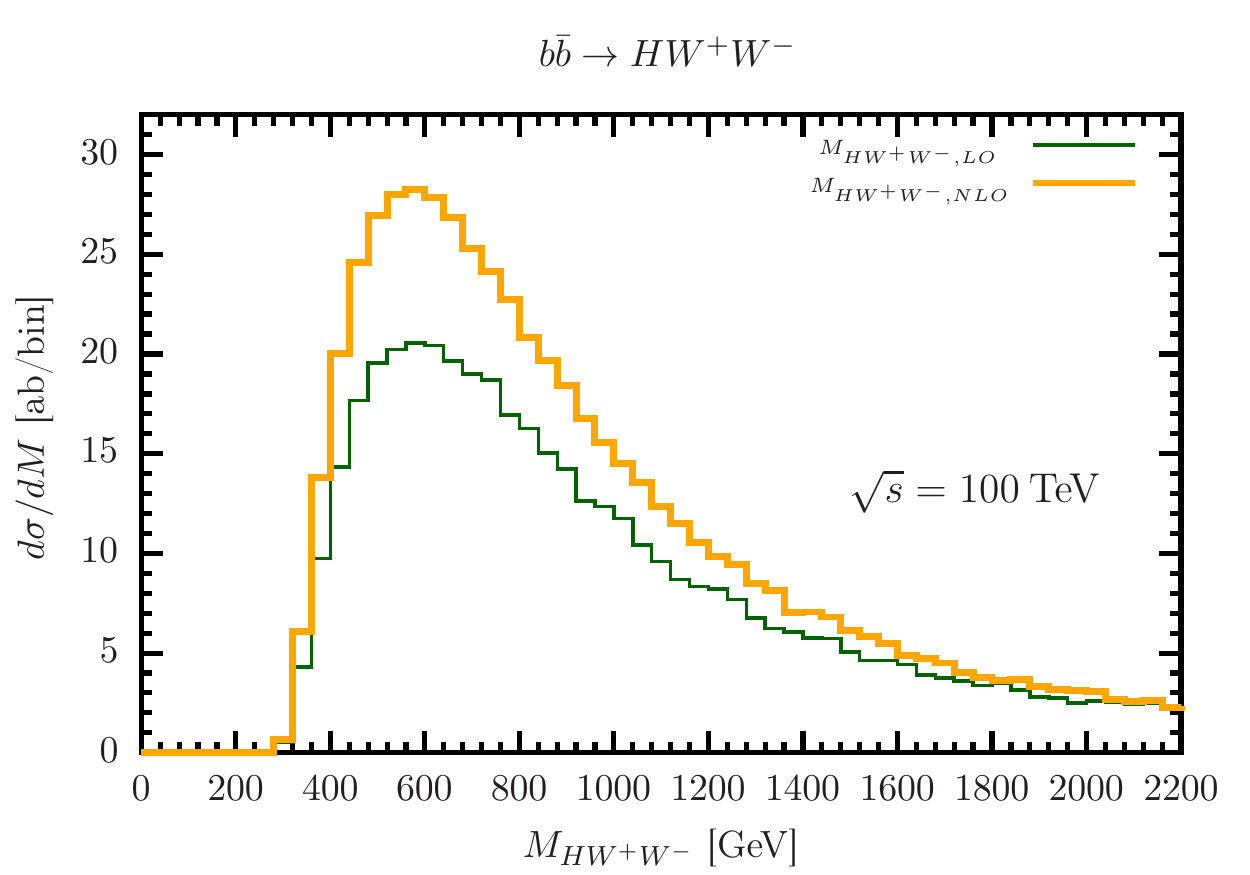}
\caption{The LO and NLO differential cross section distribution with respect to transverse momentums ($p_T$) and invariant masses ($M_{ij/ijk}$) for $100$ TeV CME.} 
\label{fig:bb2wwh_LO_NLO_pt_inv_mas}
\end{figure}

We have plotted a few different kinematical distributions at the NLO level in Fig.~\ref{fig:bb2wwh_NLO_pt_inv_mas} and Fig.~\ref{fig:bb2wwh_NLO_eta_cos0}. In Fig.~\ref{fig:bb2wwh_NLO_pt_inv_mas}, the upper-panel histograms are for the transverse momentum($p_T$) of final state particles at $14$ and $100$ TeV CMEs. As expected $W$ bosons $p_T$ distributions almost coincide with each other. The $p_T$ distributions of the Higgs boson is a bit harder. The differential cross sections are maximum around $p_T=100$ TeV for the Higgs boson and near $p_T=80$ TeV for the $W$ bosons. In the lower panel of Fig.~\ref{fig:bb2wwh_NLO_pt_inv_mas}, we have plotted the histograms for the different invariant masses ($M_{ij,ijk}$) at $14$ and $100$ TeV CMEs. Invariant mass thresholds are around $\sim210$, $\sim170$, $\sim290$ TeV and distributions are peaked around $250$, $230$, $490$ TeV for $M_{HW}$, $M_{WW}$ and $M_{HWW}$ respectively. In Fig.~\ref{fig:bb2wwh_NLO_eta_cos0}, we have plotted differential cross sections with respect to rapidity ($\eta$) of final state particles and cosine angle ($\cos\theta$) between the two final state particles for $100$ TeV CME. The distributions have maxima around $\eta=$ $0$, $-0.4$ and $0.4$ for the Higgs boson, $W^+$ and $W^-$ boson respectively. From the $\cos\theta$ plot in Fig.~\ref{fig:bb2wwh_NLO_eta_cos0}, it is clear that maximum contributions come when two final state particles are near to collinear region i.e, $\theta\sim0\:\rm{or}\:\pi$. In Fig.~\ref{fig:bb2wwh_LO_NLO_pt_inv_mas}, we have plotted the LO and the NLO  distributions to show the effect of the one-loop QCD corrections. The distributions are for only 100 TeV CME. The behavior for the 14 TeV CME is similar.
In the upper half of Fig.~\ref{fig:bb2wwh_LO_NLO_pt_inv_mas}, $p_T$ distributions are plotted and in the lower half of Fig.~\ref{fig:bb2wwh_LO_NLO_pt_inv_mas}, invariant masses have been plotted at $100$ TeV CME. We see a increase for the smaller values of the kinematic variables in all the plotted distributions.
\subsection{Anomalous coupling effect}
\label{subsec:numr_res_bsm}
As we discussed in the introduction, $VVHH$ coupling in SM is only loosely bound so far. We allow $WWHH$ coupling to deviate from the SM value in the search for new physics in the context of $\kappa$-framework~\cite{LHCHiggsCrossSectionWorkingGroup:2012nn,Ghezzi:2015vva}. In the $\kappa$ framework, only SM couplings deviate by a scale factor. $\kappa$ is defined as the deviation from the SM coupling. It is a scale factor. Although $WWH$ and $WWHH$ couplings in the SM are related but in many Effective Field Theory frameworks, these couplings can vary independently~\cite{Bishara_2017}. As there is no QCD correction to $WWHH$-vertex, the anomalous coupling will not affect the renormalization.  We have checked that the UV and IR poles cancel with the same CTs and dipole terms as in the SM. We denote deviation of $VVHH$ coupling from the SM as $\kappa_{V_2H_2}$ and $\kappa_{V_2H_2}=1$ in the SM. In this framework, we vary $\kappa_{V_2H_2}$ from $-2.0$ to $2.0$ and calculate the relative increment \big($\rm {RI}=\frac{\sigma_{\kappa}-\sigma_{SM}}{\sigma_{SM}}$\big) in the total cross section, whereas the $\kappa$ for other SM couplings are set to $1$. We choose $\kappa_{V_2H_2}=-2.0,\:-1.0,\:1.5,\:2.0$ and tabulate the results for the LO and NLO cross sections at $14$ and $100$ TeV CMEs in Table~\ref{table:bb2WWH_BSM}. It is clear from Table~\ref{table:bb2WWH_BSM} that cross sections are lower than SM prediction when $\kappa_{V_2H_2}$ is positive and higher than the SM predictions when $\kappa_{V_2H_2}$ is negative. There is not a significant relative increment ($-0.8\rm {\:to\:}+6.7\%$) at $14$ TeV. At $100$ TeV, relative increment vary from $-5.6\%$ to $+27.8\%$ for the LO cross section and from $-7.0\%$ to $+26.2\%$ for the NLO cross section. There is also $HHH$ coupling involved in this process. We also observe the $HHH$ anomalous coupling effect on the total cross sections. We vary corresponding $\kappa_{H_3}$ from $0.5$ to $2.0$. We see that there is no significant change in the LO as well as the NLO cross sections and relative increase are smaller than $1\%$ for $14$ and $100$ TeV CMEs. We see something very interesting in 
Table~\ref{table:bb2WWH_BSM_00}. The cross sections for the two longitudinally polarized $W$ bosons configuration have stronger dependence on the $\kappa_{V_2H_2}$. For the NLO cross sections the dependence is almost twice as strong as in the total cross sections. This again demonstrates the importance of measuring the polarization of the W bosons. However this dependence is weaker as compared to the LO cross sections. The difference in this dependence
underlines the importance of considering the NLO corrections.
\begin{table}[H]
\begin{center}
\begin{tabular}{|c|c|c|c|}
\hline
CME(TeV)&$\kappa_{V_2H_2}$&$\sigma^{LO}$[ab]\quad\quad RI\quad\quad\quad&$\sigma^{NLO}$[ab]\quad\quad RI\quad\quad\quad\\
\hline
\multirow{5}{*}{$14$}
&$1.0$ (SM)&$252\quad\quad\quad\quad\quad\quad$&$300\quad\quad\quad\quad\quad\quad$\\
\cline{2-4}
&$\phantom{-}2.0$&$250$\:\:\quad\quad$[-0.8\%]$&$297$\:\:\quad\quad$[-1.0\%]$\\
&$\phantom{-}1.5$&$250$\:\:\quad\quad$[-0.8\%]$&$298$\:\:\quad\quad$[-0.7\%]$\\
&$-1.0$&$262$\:\:\quad\quad$[+4.0\%]$&$310$\quad\quad\:\:$[+3.3\%]$\\
&$-2.0$&$269$\:\:\quad\quad$[+6.7\%]$&$318$\:\:\quad\quad$[+6.0\%]$\\
\hline
\multirow{5}{*}{$100$}
&$1.0$(SM)&$20671\quad\quad\quad\quad\quad\quad\:\:\:$&$27221\quad\quad\quad\quad\quad\quad$\\
\cline{2-4}
&$\phantom{-}2.0$&$19520$\:\:\quad\quad$[-5.6\%]\quad$&$25313$\:\:\quad\quad$[-7.0\%]$\\
&$\phantom{-}1.5$&$20025$\:\:\quad\quad$[-3.1\%]\quad$&$26280$\:\:\quad\quad$[-3.5\%]$\\
&$-1.0$&$24167$\:\quad\quad$[+16.9\%]\quad$&$30849$\quad\quad$[+13.3\%]$\\
&$-2.0$&$26417$\:\quad\quad$[+27.8\%]\quad$&$34356$\quad\quad$[+26.2\%]$\\
\hline
\end{tabular}
\caption{Effect of anomalous $WWHH$ coupling on the LO and NLO cross sections at the $14$ and $100$ TeV CMEs.}
\label{table:bb2WWH_BSM}
\end{center}
\end{table}

\begin{table}
\begin{center}
\begin{tabular}{|c|c|c|c|}
\hline

$\kappa_{V_2H_2}$&$\sigma^{LO}$[ab]\quad\quad RI\quad\quad\quad&$\sigma^{NLO}$[ab]\quad\quad RI\quad\quad\quad\\
\hline
$1.0$ (SM)&$7760\quad\:\:\quad\quad\quad\quad\quad$&$13610\:\:\quad\quad\quad\quad\quad\quad$ \\
$\phantom{-}2.0$&$6626$\:\:\quad\quad$[-14.6\%]$&$11850$\:\:\quad\quad$[-12.9\%]$\\
$\phantom{-}1.5$&$7173$\:\:\:\quad\quad$[-7.5\%]$\:&$12552$\quad\quad\:\:$[-7.7\%]$\:\:\\
$-1.0$&$11193$\:\quad\quad$[+44.2\%]$&$17448$\quad\quad\:\:$[+28.2\%]$\\
$-2.0$&$13416$\:\quad\quad$[+72.9\%]$&$20790$\:\:\quad\quad$[+52.8\%]$\\
\hline
\end{tabular}
\caption{Effect of anomalous $VVHH$ coupling on `00' mode at $100$ TeV CME.}
\label{table:bb2WWH_BSM_00}
\end{center}
\end{table}

\begin{figure}[!hbt]
\includegraphics [angle=0,width=0.5\linewidth]{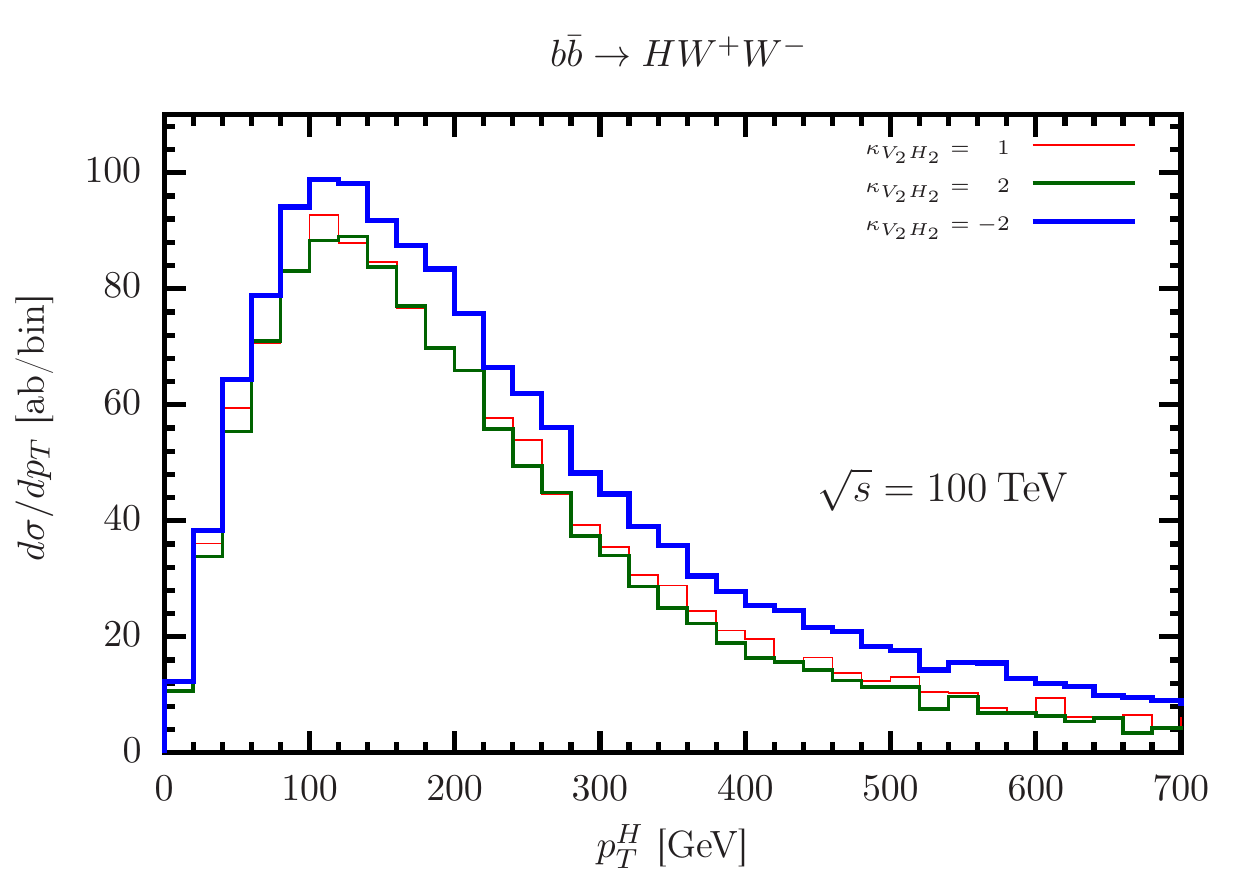}
\includegraphics [angle=0,width=0.5\linewidth]{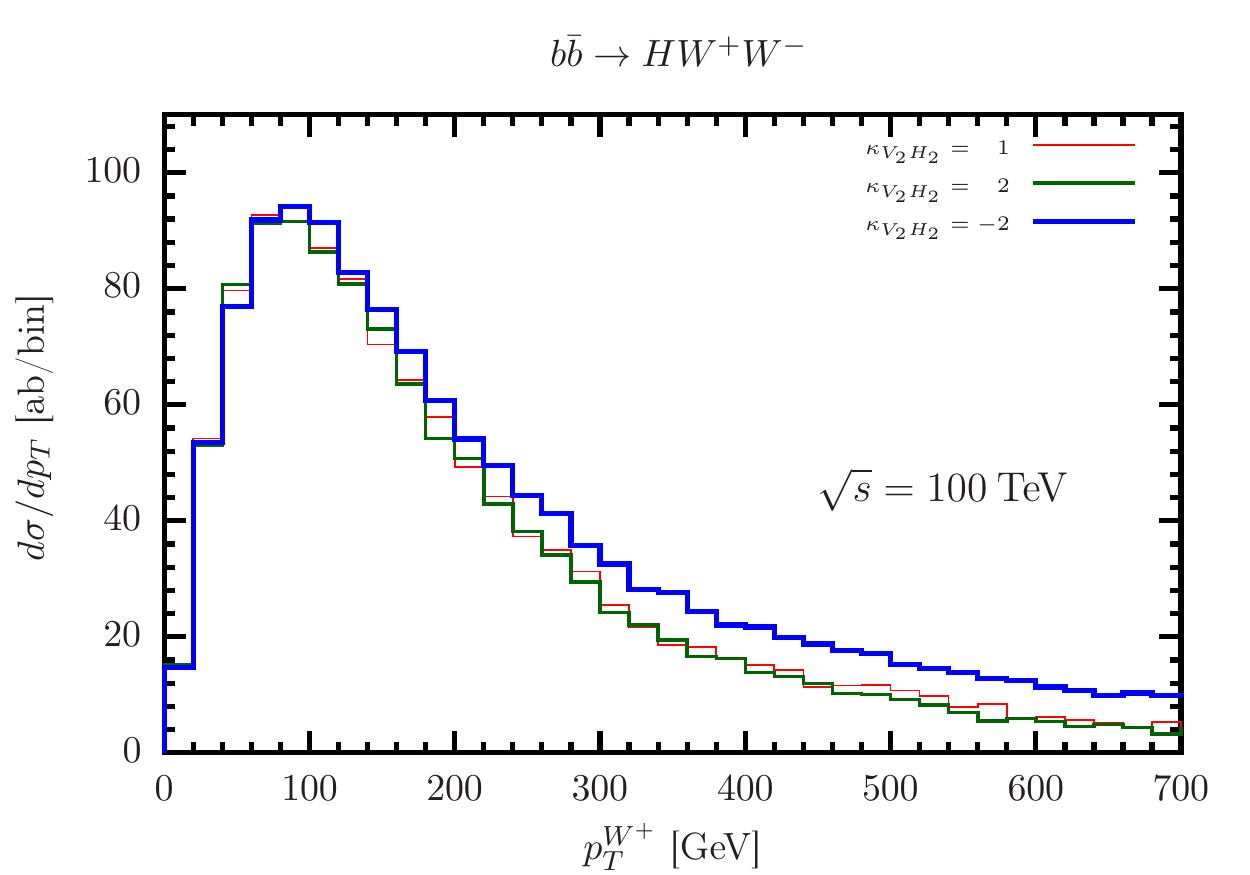}
\includegraphics [angle=0,width=0.5\linewidth]{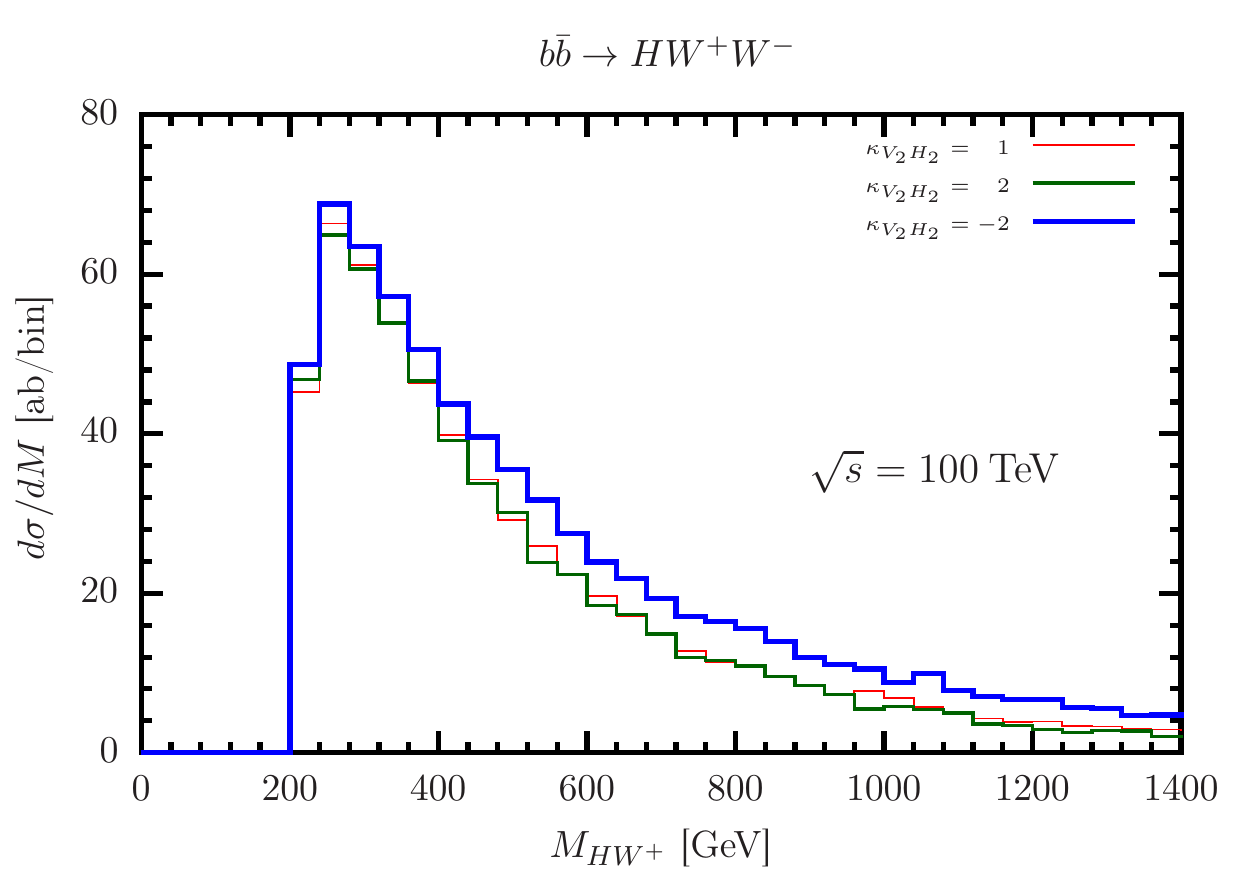}
\includegraphics [angle=0,width=0.5\linewidth]{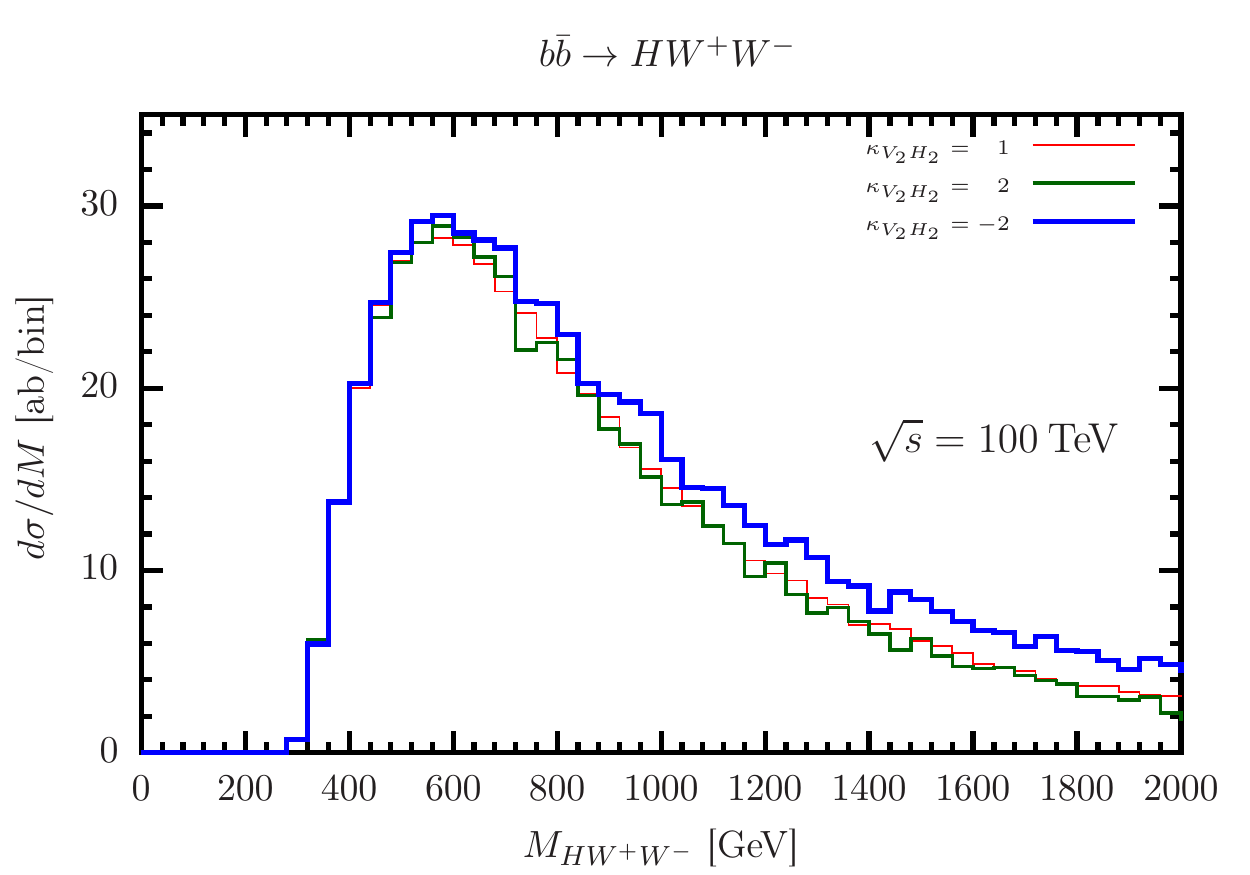}
\caption{Effect of anomalous $VVHH$ coupling on differential cross section distribution at $100$ TeV CME. Upper panel plots are for the transverse momentum of Higgs boson($p^H_T$) and $W^+$ boson ($p_T^{W^+}$). Lower panel plots are for the  $H\textendash W^+$      ($M_{HW^+}$) and $H\textendash W^+\textendash W^-$ ($M_{HW^+W^-}$) invariant masses.}  
\label{fig:bb2wwh_BSM_NLO_pt_inv_mas}
\end{figure}

In Fig.~\ref{fig:bb2wwh_BSM_NLO_pt_inv_mas}, we have plotted the NLO differential cross section distributions for the Higgs boson and $W^+$ boson transverse momenta,
and different invariant masses. The maxima of the differential cross sections are 
about at the same value  as for the SM. As there is not that much increase for $\kappa_{V_2H_2}=2$, the corresponding distributions nearly overlap with the SM. On the other hand, we see a sharp deviation in distributions from the SM for $\kappa_{V_2H_2}=-2$. Interesting fact about the negative $\kappa_{V_2H_2}$ is that
the distribution are harder. This difference in the shape can be used in putting a strong bound on the coupling. One could put a cut on $p_T^{W}$, or one of the plotted
invariant masses to select events with a larger component of anomalous events.
\section{Conclusion}
\label{sec:conclusion}
    In this letter, we have focused on the NLO QCD corrections to $b {\bar b} \to WWH$. This process has 
	significant dependence on $WWHH$ coupling. But, the contribution of this process to $pp \to WWH$
	is only about $15-20\%$ of that of light quark scattering. 
This is where the consideration of the polarization of  the W bosons helps. 
When  both the $W$ bosons are longitudinally polarized, then this fraction can increase to $70-80\%$.
 It turns out that the NLO QCD corrections are
also largest for this polarization configuration, making the
dependence on the $WWHH$ coupling even stronger.
For example, at the 100 TeV CME, the NLO corrections are about $32\%$, but the corrections are about $75\%$, when both final state $W$ bosons are longitudinally polarized.
Our study suggests that the measurement of the polarization of the final state
$W/Z$ bosons can be a useful tool to measure the
couplings of the vector bosons and Higgs boson. We have also examined the
effect of the variation of $\kappa_{V_2H_2}$. The variation in the cross section
can be twice as large when we consider longitudinally polarized $W$ bosons. In addition, we find that the invariant mass and the $p_T^{W}$ distributions are considerably harder for the negative values of $\kappa_{V_2H_2}$. This can also be useful to put a stronger bound on the coupling. However, to find the bound, one would need to do a detailed background analysis which we leave for the future.

 \section*{Acknowledgements}
PA would like to acknowledge fruitful discussions with Debashis Saha and Ambresh Shivaji. Part of this work was done when PA was visiting IIT, Delhi. BD would like to acknowledge the useful discussions with Debashis Saha.  

%
%
%
%

\bibliographystyle{JHEP}
\bibliography{bb2wwhref}
\end{document}